\documentclass[lettersize,journal]{IEEEtran}
\usepackage{amsmath,amsfonts}
\usepackage{algorithmic}
\usepackage{array}
\usepackage[caption=false,font=normalsize,labelfont=sf,textfont=sf]{subfig}
\usepackage{textcomp}
\usepackage{stfloats}
\usepackage{url}
\usepackage{verbatim}
\usepackage{graphicx}

\usepackage{booktabs}
\usepackage{multirow}

\def\BibTeX{{\rm B\kern-.05em{\sc i\kern-.025em b}\kern-.08em
    T\kern-.1667em\lower.7ex\hbox{E}\kern-.125emX}}
\usepackage{balance}
\begin{document}
\title{An Efficient FPGA-Based Accelerator for Swin Transformer}
\author{Zhiyang Liu$^{1}$, Zhenhua Ren$^{1}$, Pengyu Yin$^{2}$
\thanks{$^{1}$Authors are with the School of Automation Engineering, University of Electronic Science and Technology of China(UESTC), Chengdu, China. {\tt\small\{202121060140, 202221060711\}@std.uestc.edu.cn;}} 
\thanks{$^{2}$Authors are with the Centre for Advanced Robotics Technology Innovation (CARTIN), School of Electrical and Electronic Engineering, Nanyang Technological University, Singapore. 
{\tt\small\{pengyu001\}@e.ntu.edu.sg;}}
}


\maketitle

\begin{abstract}
Since introduced, Swin Transformer has achieved remarkable results in the field of computer vision, it has sparked the need for dedicated hardware accelerators, specifically catering to edge computing demands. For the advantages of flexibility, low power consumption, FPGAs have been widely employed to accelerate the inference of convolutional neural networks (CNNs) and show potential in Transformer-based models. Unlike CNNs, which mainly involve multiply and accumulate (MAC) operations, Transformer involve non-linear computations such as Layer Normalization (LN), Softmax, and GELU. These nonlinear computations do pose challenges for accelerator design. In this paper, to propose an efficient FPGA-based hardware accelerator for Swin Transformer, we focused on using different strategies to deal with these nonlinear calculations and efficiently handling MAC computations to achieve the best acceleration results. We replaced LN with BN, Given that Batch Normalization (BN) can be fused with linear layers during inference to optimize inference efficiency. The modified Swin-T, Swin-S, and Swin-B respectively achieved Top-1 accuracy rates of 80.7$\%$, 82.7$\%$, and 82.8$\%$ in ImageNet. Furthermore, We employed strategies for approximate computation to design hardware-friendly architectures for Softmax and GELU computations. We also designed an efficient Matrix Multiplication Unit to handle all linear computations in Swin Transformer. As a conclude, compared with CPU (AMD Ryzen 5700X), our accelerator achieved 1.76x, 1.66x, and 1.25x speedup and achieved 20.45x, 18.60x, and 14.63x energy efficiency (FPS/power consumption) improvement on Swin-T, Swin-S, and Swin-B models, respectively. Compared to GPU (Nvidia RTX 2080 Ti), we achieved 5.05x, 4.42x, and 3.00x energy efficiency improvement respectively. As far as we know, the accelerator we proposed is the fastest FPGA-based accelerator for Swin Transformer.
\end{abstract}


\section{Introduction}\label{Introduction}
\IEEEPARstart{T}{ransformer,} proposed by Vaswani et al. in 2017\cite{vaswani2017attention}, is a revolutionary neural network model based on self-attention mechanisms that achieved significant success in the field of Natural Language Processing (NLP). Inspired by this, researchers sought to apply Transformer-like self-attention mechanisms to the field of computer vision, leading to groundbreaking work.

Some models based on the Transformer architecture, such as VIT\cite{dosovitskiy2020image}, TNT\cite{han2021transformer}, and DeiT\cite{touvron2021training}, segment images into a sequence of image patches and process them using self-attention mechanisms. This enables these models to better comprehend the semantic information within an image, overcoming the limitations imposed by local perceptions. As a result, they have achieved remarkable success in computer vision tasks.

Liu et al. \cite{liu2021swin} have proposed a novel visual Transformer architecture, Swin Transformer. It aims to tackle the computational and memory overhead challenges faced by traditional Transformers when handling large-sized images. By employing sliding windows and a hierarchical structure, Swin Transformer has emerged as a new backbone in the field of computer vision. It has achieved state-of-the-art (SOTA) performance in various machine vision tasks, including image classification, object detection, and semantic segmentation.

For some tasks such as autonomous driving and face recognition, computer vision tasks need to be implemented in real-time at the edge. Given the impressive performance of Swin Transformer in computer vision tasks, designing a hardware accelerator tailored for Swin Transformer in edge computing is indeed necessary.

FPGA (Field-Programmable Gate Array) is an integrated circuit that can be programmed to implement specific functions. Their characteristics such as parallel computing capabilities, low power consumption and low latency make FPGAs highly promising as neural network accelerators for enhancing computational performance, reducing power consumption, and achieving low latency. Additionally, the theoretical foundations and optimization methods for designing neural network accelerators like CNNs based on FPGA have become well-established \cite{zhang2015optimizing}, \cite{ma2017optimizing}, \cite{ma2018optimizing}. 

Unlike CNNs, the Transformer not only involves a significant amount of matrix multiplication but also extensively employs nonlinear functions like Softmax and the GELU\cite{hendrycks2016gaussian} activation function. These nonlinear functions heavily rely on exponential, Hyperbolic tangent (tanh) calculation, and division operations, which not only consume a substantial amount of FPGA resources such as DSP and LUT but also introduce increased runtime latency. With fully studying about the formulas of Softmax and GELU, we can transform both two functions above into computations involving relatively hardware-friendly base-2 exponentiation and division operations, and design achitectures to approximate execute above operations utilizing shift, add addition, multiplication operations.

Besides non-linear fuctions, Swin Transformer also consist a significant amount of complex linear computations, including convolutions and matrix multiplications. These linear computations indeed constitute the vast majority of the model's computational workload. In \cite{wang2022via}, \cite{nag2023vita}, \cite{hu2022hardware}, these works do not employ a single computation engine architecture and lack analysis based on the characteristics of convolutions in Swin Transformer, as well as the shapes of matrices involved in matrix multiplications, made it hardly exploit the computition efficiency.

Moreover, in the case of Swin Transformer, the extensive use of Batch Normalization(BN) \cite{ioffe2015batch} in CNNs has been replaced with Layer normalization(LN) \cite{ba2016layer} for normalization purposes. BN can precompute running means and variances during inference, and the fusion of BN with convolutional layers is a commonly employed hardware acceleration technique \cite{jaderberg2014speeding}. However, LN requires real-time computation of input mean and variance. The division and square root calculations involved in LN computations are also not well-suited for hardware like FPGA due to their computational complexity. 

In this paper, based on the characteristics of convolutions in Swin Transformer, we transform the convolution in PatchEmbed into matrix multiplications, while employing a single computation engine architecture to compute all matrix multiplications. Meanwhile, inspired by \cite{shen2020powernorm}, \cite{yao2021leveraging}, \cite{chen2021empirical} we replicated and validated the scheme of using BN instead of LN in Swin-T, Swin-S, and Swin-B models. Ultimately, we achieved top-1 accuracy rates of 80.7$\%$, 82.7$\%$, and 82.8$\%$ on ImageNet, respectively. The decrease in accuracy compared to using LN normalization was 0.6$\%$, 0.3$\%$, and 0.7$\%$, respectively, which we deemed acceptable for our purposes.

In conclusion, we proposed an efficient FPGA-based accelerator tailored for the Swin Transformer, employed strategies to design hardware-friendly architectures for efficient model inference. The main contributions of our work are summarized as follows:

\begin{enumerate}{}{}
\item{We replaced LN with BN in the Swin Transformer, verified the feasibility of this strategy, and fused BN with linear layers for efficient inference.}
\item{We designed a novel efficient FPGA Matrix Multiplication Unit (MMU) and related accelerator optimizations for all Swin Transformer Linear Computing to achieve the best multiply and accumulate (MAC) handle performance.}
\item{We propose hardware-friendly 16-bit fixed architectures for efficiently computing Softmax and GELU functions.}
\item{We designed accelerators based on Xilinx XCZU19EG FPGA for Swin-T, Swin-S, and Swin-B respectively. These accelerators respectively achieve 1.76x, 1.66x, and 1.25x speedup and 20.45x, 18.60x, and 14.63x energy efficiency compared to the inferences in AMD Ryzen 5700X CPU, also achieve 0.20x, 0.17x, and 0.12x speedup and 5.05x, 4.42x and 3.00x energy efficiency compared to a high-end GPU albeit at the expense of the inference speed.}
\end{enumerate}

The remaining sections of this paper are structured as follows:

Section 2 serves as a research background, providing an in-depth overview of the detailed principles and structure of the Swin Transformer. It also covers background about various non-linear operations, including Layer Normalization and Batch Normalization, Softmax, as well as the GELU activation function. In addition, we have provided an overview of existing research on FPGA hardware accelerators.

Section 3, as a theoretical foundation of the research motivation, is divided into two parts. The first part elucidates the motivation behind our replacement of the LN scheme with the BN scheme in Swin Transformer. It describes the modifications to the Swin Transformer structure after adopting the BN scheme, along with the approach of fusing BN with subsequent linear operation layers during inference. The second part delves into the approximation computation methods for Softmax and GELU functions.

Section 4 presents the architecture of the accelerator, encompassing both the overall structure of the accelerator and its dataflow, as well as detailing the structure and operational principles of its submodules.

Section 5 presents the results of the experiment, including a comparison with previous related work and a comparison with CPU and GPU.

\section{BACKGROUND}\label{BACKGROUND}
\subsection{Swin Transformer}

\begin{figure*}[!t]
\centering
\includegraphics[width=6.5in]{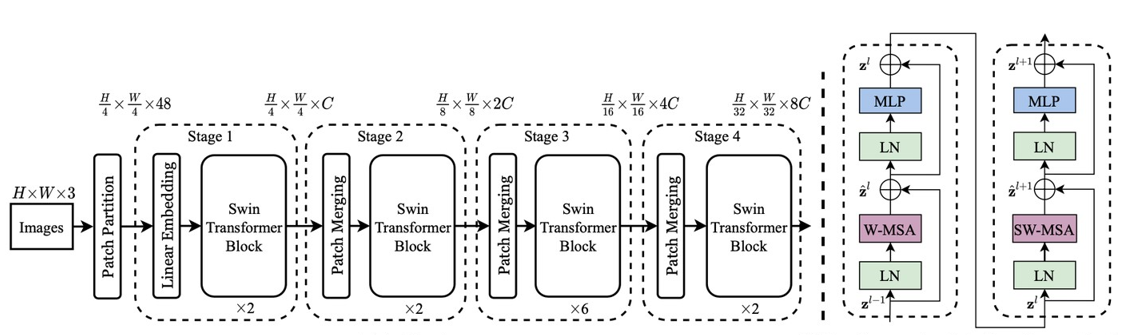}
\caption{The overall architecture of the Swin Transformer.}
\label{Swin Transformer}
\end{figure*}

Swin Transformer is an image classification model based on the Transformer architecture. Its principle involves applying the concept of self-attention to computer vision tasks. The design inspiration for the Swin Transformer derives from both traditional attention mechanisms and hierarchical feature extraction concepts while incorporating innovative improvements at a detailed level. The structure of the Swin Transformer is shown in Fig. \ref{Swin Transformer}.

\textbf{Window Self-Attention} is a critical concept in the Swin Transformer model, serving as one of its core components. It decomposes attention computation into multiple local windows, conducting self-attention calculations within each window. This decomposition enables parallel computations within each window, thereby reducing the computational burden. This design, while maintaining model performance, significantly decreases computational complexity and memory overhead, making it particularly well-suited for handling high-resolution images.

To facilitate improved interaction with other windows, the Swin Transformer also introduces the shifted window operation. This alignment enables the local window to more accurately capture crucial contextual information, enhancing the model's expressive capability. Through appropriate offset strategies, the Shifted Window Attention can also reduce overlap between local windows, further lowering computational complexity and enhancing computational efficiency.

Simultaneously, the Swin Transformer employs a hierarchical design through Patch Merging downsampling layers, gradually expanding the receptive field. This expansion enables the attention mechanism to encompass global features, enhancing the model's ability to capture broader context.

\subsection{Non-Linear Functions}

\subsubsection{Layer Normalization(LN) and Batch Normalization(BN)}

LN and BN are two common normalization techniques in neural networks. BN \cite{ioffe2015batch} is a technique for normalizing inputs at each layer of a neural network and is widely used in vision tasks in convolutional neural networks (CNNs). It normalizes each batch of samples to maintain relatively stable inputs for each layer.

Unlike BN, LN \cite{ba2016layer} normalizes each sample's features at each layer, rather than across batches. It is widely adopted in NLP tasks in Transformer networks. Simultaneously, LN is widely employed in visual Transformer models, including Swin Transformer.

\subsubsection{Softmax  Function}

In the Self-Attention mechanism, the Softmax operator is utilized to compute attention weights, which are then normalized into a probability distribution. This normalization enables a weighted summation of inputs from various positions. Such weighted summation allows the model to emphasize different positions of information based on their relationships in the input. The Softmax function is defined by the following equation:

\begin{equation}
\label{deqn_ex1}
f\left(x_i\right)=\frac{e^{x_i}}{\sum_{j=1}^N e^{x_j}}(i=1,2, \ldots, N)
\end{equation}

\subsubsection{GELU (Gaussian Error Linear Unit)}

GELU is an activation function widely employed in Transformers, including Swin Transformer. In Swin Transformer, the GELU activation function is typically utilized for the non-linear transformations in the Feed-Forward Networks (FFN), used to perform non-linear transformations on the outputs of the self-attention mechanism. In comparison to the commonly used ReLU activation function in CNNs, the computation of the GELU function involves complex operations such as multiplication, Hyperbolic tangent (tanh) calculation, and cubing. While GELU performs well in enhancing model performance, its computational complexity is relatively high. This underscores the importance of optimization and design on hardware platforms.

The various components within the architecture of Swin Transformer utilize multiple nonlinear functions to enhance the model's representational capacity and learning capabilities. Designing FPGA accelerators to support the computation of non-linear functions such as LN, Softmax, and GELU within the model is a complex task that demands in-depth hardware design and optimization. This endeavor stands as a critical aspect of deploying Transformer models onto hardware platforms.

\subsection{FPGA-Based Accelerators}

FPGAs (Field-Programmable Gate Arrays) have been widely applied in accelerating neural networks, particularly in the domain of CNNs. In CNN accelerators, based on the parallel computing capability and hardware customization of FPGAs. Researchers in \cite{zhang2015optimizing}, \cite{ma2017optimizing}, \cite{ma2018optimizing} have extensively optimized loop methodologies and dataflow in accelerators to maximize FPGA resource utilization and to get the best performance. Furthermore, to overcome the limitations of FPGA computing capabilities, researchers in \cite{liang2019evaluating} have employed Winograd and FFT algorithms to design efficient CNN hardware accelerators. Their research points out the direction for the design of hardware accelerators. 

Due to the computational complexity and complex
model structures of transformers, there is a relatively limited amount of work focused on FPGA hardware accelerators designed for transformers. Furthermore, most of these accelerators \cite{li2020ftrans}, \cite{peng2021accelerating} are predominantly applicable to the field of Natural Language Processing (NLP). For visual transformer accelerators, \cite{wang2022via}, \cite{lit2022auto}, \cite{zhao2022fpga} have respectively proposed FPGA accelerators and optimization methods for Swin-T, DEiT, and TNT. However, these researches still lack hardware optimizations for non-linear functions such as LN, softmax, and GELU in the context of visual transformer accelerators. This provides us with the motivation to delve into the development of FPGA accelerators tailored for visual transformers.

\section{MOTIVATION}\label{MOTIVATION}
\subsection{Raplace Layer Normalization with Batch Normalization}

In the inference process of CNNs, merging BN into adjacent convolutional layers is a mature technique that can significantly enhance inference speed. However, in Swin Transformer, LN requires real-time computation of the mean and variance of the last dimension's input. This characteristic prevents layer normalization from being seamlessly integrated into convolutional or linear computation layers, similar to BN. Moreover, LN involves division and square root operations, which incur substantial hardware resource overhead in FPGAs and can be cumbersome to implement. 

Some research studies \cite{kim2021bert, wei2021floating, andraka1998survey} attempted to use methods such as Newton's iterative method, Taylor series expansion, and the CORDIC algorithm to approximate division and square root calculations. Although these methods are more hardware-friendly, compared to the ability to fuse BN into linear computation layers, the computation process of LN not only increases hardware resource overhead but also introduces additional latency. BN is faster in inference than LN due to the avoidance of calculating the mean and variance statistics during inference.

Thus, we had the idea of replacing layer normalization with batch normalization in Swin Transformer.

Shen et al. \cite{shen2020powernorm} investigated four statistics during training with Batch Normalization in Transformers: batch mean and variance, as well as the mean and variance of their gradients. The batch mean and variance remained relatively stable in computer vision tasks. As for the gradients' mean and variance, they exhibited minimal oscillation in computer vision tasks, and no outliers were observed after training completion. Which demonstrates the potential of utilizing BN in Swin Transformer.

Furthermore, \cite{yao2021leveraging} conducted an exhaustive validation and analysis of substituting BN for LN in Transformer-based vision architectures. It was pointed out that directly replacing the LN in Transformer with BN can lead to non-convergence during training and even model instability or collapse. The paper thoroughly analyzed the issue of model frequent crashes in training resulting from LN-to-BN replacement and proposed a solution to address the training instability: introducing an additional layer of BN in between the two linear layers of the FFN. After making modifications to the model, the paper trained the DeiT-Small, Swin-based, and SwinD-based models. The results indicated that all models experienced a decrease in accuracy within a range of 1$\%$.

\begin{figure}[!t]
\centering
\includegraphics[width=2.5in]{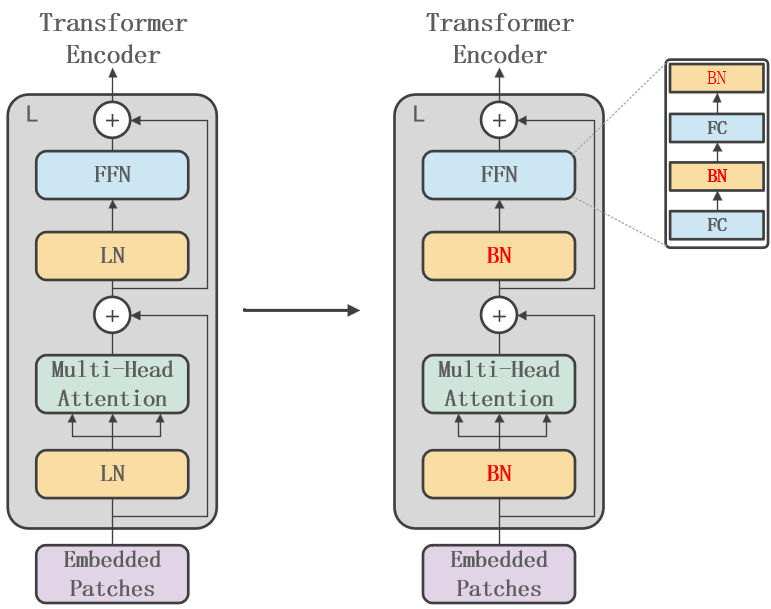}
\caption{The modified transformer block structure.}
\label{BN}
\end{figure}

Based on the aforementioned research findings, we are prepared to replicate the accomplishments of \cite{yao2021leveraging}. We are planning to undertake a thorough validation by attempting to replace all instances of Layer Normalization with Batch Normalization in Swin-T, Swin-S, and Swin-B models, respectively. Additionally, we will add two additional batch normalization layers after the two linear layers in the FFN. The modified transformer block structure is shown in Fig. \ref{BN}. The revamped model has achieved promising training outcomes. For further details, please refer to Section Five.

In CNNs, it is common practice to fuse batch normalization into the preceding convolutional layer. Due to the presence of the Shortcut Mechanism (residual connections) in the Transformer encoder, BN cannot be directly fused with the preceding layer as is done in convolutional neural networks. Residual connections involve adding the output of the previous layer to the subsequent layer, and BN can disrupt the nature of these residual connections, making it challenging to straightforwardly fuse with the previous layer. As a result, in the Transformer encoder, special methods or techniques are often required to handle the application of BN while preserving the effectiveness of residual connections. Our approach is to fuse BN into layers of linear computation.

 During the inference process, freezed BN can be regarded as a 1x1 convolution operation. In this perspective, the formula for BN can be expressed as follows:

\begin{scriptsize} 
\begin{equation}
\begin{aligned}
\label{deqn_ex2}
\left(\begin{array}{c}
\hat{F}_{1, i, j}\\
\hat{\boldsymbol{F}}_{2, i, j}\\
\vdots\\
\hat{\boldsymbol{F}}_{C-1, i, j}\\
\hat{\boldsymbol{F}}_{C, i, j}
\end{array}\right)=\left(\begin{array}{ccccc}
\frac{\gamma_1}{\sqrt{\hat{\sigma}_1^2+\epsilon}} & 0 & \cdots & & 0\\
0 & \frac{\gamma_2}{\sqrt{\hat{\sigma}_2^2+\epsilon}} & & &\\
\vdots & & \ddots & &\\
& & & \frac{\gamma_{C-1}}{\sqrt{\hat{\sigma}_{C-1}^2+\epsilon}} & 0\\
0 & & \ldots & 0 & \frac{\gamma_C}{\sqrt{\hat{\sigma}_C^2+\epsilon}}
\end{array}\right)
\\\times\left(\begin{array}{c}
F_{1, i, j}\\
F_{2, i, j}\\
\vdots\\
F_{C-1, i, j}\\
F_{C, i, j}
\end{array}\right)+\left(\begin{array}{c}
\beta_1-\gamma_1 \frac{\widehat{\mu}_1}{\sqrt{\widehat{\sigma}_1^2+\varepsilon}}\\
\beta_2-\gamma_2 \frac{\widehat{\mu}_2}{\sqrt{\widehat{\sigma}_2^2+\varepsilon}}\\
\vdots\\
\beta_{C-1}-\gamma_{C-1} \frac{\hat{\mu}_{C-1}}{\sqrt{\widehat{\sigma}_{C-1}^2+\varepsilon}}\\
\beta_C-\gamma_C \frac{\widehat{\mu}_C}{\sqrt{\widehat{\sigma}_C^2+\varepsilon}}
\end{array}\right)
\end{aligned}
\end{equation}
\end{scriptsize}

For a feature map $F$, the normalized result is $\hat{F}$, the BN layer's freezed parameters noted as $\hat{\sigma}^2$, $\hat{\mu}$, $\gamma$, $\beta\in\mathbb{R}^{\mathit{C}}$.

Matrix multiplication is also regarded as a 1x1 convolutional operation. Therefore, we can represent BN and the linear layer as follows:
\begin{equation}
\label{deqn_ex3}
\hat{f}_{i, j}=\boldsymbol{W}_{\text {Linear }} \cdot\left(\boldsymbol{W}_{B N} \cdot f_{i, j}+\boldsymbol{b}_{B N}\right)+\boldsymbol{b}_{\text {Linear }}\\
\end{equation}

Where $\boldsymbol{W}_{B N} \in \mathbb{R}^{C \times C}$, $\boldsymbol{b}_{B N} \in \mathbb{R}^{\boldsymbol{c}}$ noted as the parameter of BN, and $\boldsymbol{W}_{\text {Linear }} \in \mathbb{R}^{C \times C_{\text {out }}}$, $\boldsymbol{b}_{\text {Linear }} \in \mathbb{R}^{\boldsymbol{C}_{\text {out }}}$ noted as the parameter of subsequent linear layer. As a result, the computations of BN and the linear layer can be fused into a single linear layer:
\begin{equation}
\label{deqn_ex4}
\begin{aligned}
\boldsymbol{W} & =\boldsymbol{W}_{\text {Linear }} \cdot \boldsymbol{W}_{B N} \\
\boldsymbol{b} & =\boldsymbol{W}_{\text {Linear }} \cdot \boldsymbol{b}_{B N}+\boldsymbol{b}_{\text {Linear }}
\end{aligned}
\end{equation}

The method of merging BN with the preceding linear layer is referred to as the approach in \cite{jaderberg2014speeding}.

\subsection{Approximate Calculation of Nonlinear function}

In addition to normalization operations, the Transformer encoder also includes two types of non-linear computations: Softmax and GELU.

The Softmax operation involves exponential and division computations, both of which indeed pose challenges when designing efficient hardware architectures on FPGA. Exponential and division computations require substantial computational resources and time, potentially leading to increased hardware resource consumption and computation latency. Some researchers have made contributions in this regard: \cite{vasyltsov2021efficient} proposed two methods using 8-bit fixed-point approximations based on lookup tables to compute Softmax, achieving an accuracy loss of less than 1$\%$. In \cite{zhu2020efficient}, a method utilizing lookup tables to implement a Precision-Adjustable approach for the Softmax function was employed. In \cite{khan2021npe}, a piecewise linear approach was adopted to approximate the computation of Softmax for the design of an FPGA-based BERT processor. 

GELU computation involves multiplication and hyperbolic tangent (tanh) operations, which demand significant hardware resources and time. When designing efficient hardware architectures using FPGA, challenges arise, necessitating careful consideration of how to optimize these operations to achieve high-performance and low-latency GELU computation. In order to achieve full quantization inference in Transformers, \cite{kim2021bert} proposed a method of calculating GELU using polynomial approximation. In \cite{li2023high}, the article transforms the GELU formula by converting the expression involving multiplication and hyperbolic tangent (tanh) operations into a form that includes division, multiplication, and base-2 exponentiation calculations. 

Our nonlinear function hardware architecture design is primarily inspired by \cite{vasyltsov2021efficient} and \cite{li2023high}. Firstly, it involves further transforming or approximating the Softmax and GELU functions to optimize their computations, particularly when implementing them in hardware architectures like FPGA. The objectives of these transformations or approximations are to reduce computational complexity and resource utilization while maintaining a reasonable level of accuracy.

To prevent accuracy loss caused by potential data overflow, the softmax can be represented as follows:

\begin{equation}
\label{deqn_ex5}
f\left(x_i\right)=\frac{e^{x_i-x_{\max }}}{\sum_{j=1}^n e^{x_j-x_{\max }}}, i=1,2, \ldots, n
\end{equation}

Where $x_1$, $x_2$, …, $x_n$ are inputs. $x_{max}$ is the maximum value among
$x_1$, $x_2$, …, $x_n$. 

According to the hardware architecture of an FPGA, computing exponentials with base 2 is significantly easier compared to exponentials with base $e$, and $e^{x_i}$ is equal to $2^{\log _2 e \cdot x_i}$, Therefore, we replace the $e^{x_i}$ in formulas B with $2^{\log _2 e \cdot x_i}$, and the softmax function can be transformed and represented as follows:

\begin{equation}
\label{deqn_ex6}
f\left(x_i\right)=\frac{2^{\log _2 e\left(x_i-x_{\max }\right)}}{\sum_{j=1}^n 2^{\log _2 e\left(x_j-x_{\max }\right)}}
\end{equation}

The GELU activation function can be represented as:

\begin{equation}
\label{deqn_ex7}
g\left(x_i\right)=0.5 x_i\left(1+\tanh \left(\sqrt{\frac{2}{\pi}}\left(x_i+0.044715 x_i^3\right)\right)\right)
\end{equation}

We set $h\left(x_i\right)=\sqrt{\frac{2}{\pi}}\left(x_i+0.044715 x_i^3\right)$. The GELU function can be transformed as follows:

\begin{equation}
\begin{aligned}
\label{deqn_ex8}
& g\left(x_i\right)=0.5 x_i\left(1+\frac{e^{h\left(x_i\right)}-e^{-h\left(x_i\right)}}{e^{h\left(x_i\right)}+e^{-h\left(x_i\right)}}\right) \\
& =\frac{x_i e^{h\left(x_i\right)}}{e^{h\left(x_i\right)}+e^{-h\left(x_i\right)}}=\frac{x_i}{1+2^{-2 \log _2 e \cdot h\left(x_i\right)}}
\end{aligned}
\end{equation}

Next, we need to perform FPGA hardware optimization design for formulas (6) and (8), including specialized hardware optimization for base-2 exponentiation and division operations. This aims to enhance the efficiency and speed of these calculations while minimizing resource utilization.

We further optimize the computation of the base-2 exponentiation. In formulas (6) and (8) ${log}_2e$, as the constant coefficient of the exponent of $2^{\log _2 e \cdot x_i}$, can be approximately represented in binary as: 1.0111. Where 1.0111 can be represented as 1 + 0.1 - 0.0001. Therefore, we can compute ${log}_2e\times x_i$ using two shift operations and two addition-subtraction operations. 

In formula (8), we set $s\left(x_i\right)$ equal to the exponent of $2^{-2 \log _2 e \cdot h\left(x_i\right)}$ and can be represented as: 

\begin{equation}
\begin{aligned}
\label{deqn_ex9}
s\left(x_i\right)=-2 \log _2 e \cdot \sqrt{\frac{2}{\pi}}\left(x_i+0.044715 x_i^3\right)
\end{aligned}
\end{equation}

Similarly, one of the constant coefficients $-2 \log _2 e \cdot \sqrt{\frac{2}{\pi}}$ in formula (9) can be approximately represented in binary as: -10.0101=-10-0.01-0.0001. Another constant coefficient of 0.044715 in formula (9) can be approximately represented in binary as:0.000011=0.000001+0.00001. In formula (9), the multiplication computation involving two coefficients can be accomplished using adders and shift operations.

We set $v\left(x_i\right)=\log _2 e \cdot x_i$, $v\left(x_i\right)$ can be divided into an integer part $int\left(x_i\right)$ and a fractional part $frac\left(x_i\right)$. 

The calculation of $2^{\log _2 e \cdot x_i}$ can be represented as follows(Where $<<$ denotes the shift operation.):

\begin{equation}
\begin{aligned}
\label{deqn_ex10}
2^{\log _2 e \cdot x_i} &=2^{v\left(x_i\right)}=2^{\operatorname{int}\left(x_i\right)} \cdot 2^{\operatorname{frac}\left(x_i\right)} \\
&=2^{\text {frac }\left(x_i\right)} \ll \operatorname{int}\left(x_i\right)
\end{aligned}
\end{equation}

According to the transformation of $2^{\log _2 e \cdot x_i}$ in formula (10), we convert the calculation of $2^{\log _2 e \cdot x_i}$ into the calculation of $2^{frac{\left(x_i\right)}}$ and shift operations. The input range is reduced from $-\infty<x_i<+\infty$ to $0<frac{\left(x_i\right)}<1$, facilitating the use of piecewise linear approximation to compute $2^{frac{\left(x_i\right)}}$. The detailed FPGA hardware architecture for computing  $2^{frac{\left(x_i\right)}}$ is presented in Section \ref{ACCELERATOR ARCHITECTURE}.

Furthermore, we also need to optimize division calculations in hardware. The dividend and divisor are denoted as $F_1$ and $F_2$, respectively. The dividend $F_1$ and the divisor $F_2$ can be expressed as $F_1=\ w_1\times2^{m_1}$, $F_2=\ w_2\times2^{m_2}$. Where $w_1$ and $w_2$ are integers, and $m_1,\ m_2\ \in[1,2)$. $w_1$, $w_2$, $m_1$ and $m_2$ can be easily obtained from LOD (introduced in Section Four). Thus, division can be represented as:

\begin{equation}
\label{deqn_ex11}
\frac{F_1}{F_2}=2^{\left(\log _2 m_1+w_1\right)-\left(\log _2 m_2+w_2\right)}
\end{equation}

In $[1,2)$, ${log}_2m_1$ and ${log}_2m_2$ can be respectively approximated by $m_1-1$ and $m_2-1$. Therefore, division can be further represented as:

\begin{equation}
\label{deqn_ex12}
\frac{F_1}{F_2} \approx 2^{\left(m_1-1+w_1\right)-\left(m_2-1+w_2\right)}=2^{\left(m_1+w_1\right)-\left(m_2+w_2\right)}
\end{equation}

In summary, division calculation can be transformed into exponentiation with base 2, along with shifting and addition operations.

\section{ACCELERATOR ARCHITECTURE}\label{ACCELERATOR ARCHITECTURE}

In this chapter, we will commence by providing a comprehensive overview of the accelerator's overarching architecture, coupled with an in-depth exposition of the data flow intrinsic to the accelerator's operations.  Moving on, our focus shifts to a detailed elucidation of each individual submodule's structure and functionality.  

\begin{figure*}[!t]
\centering
\includegraphics[width=6.5in]{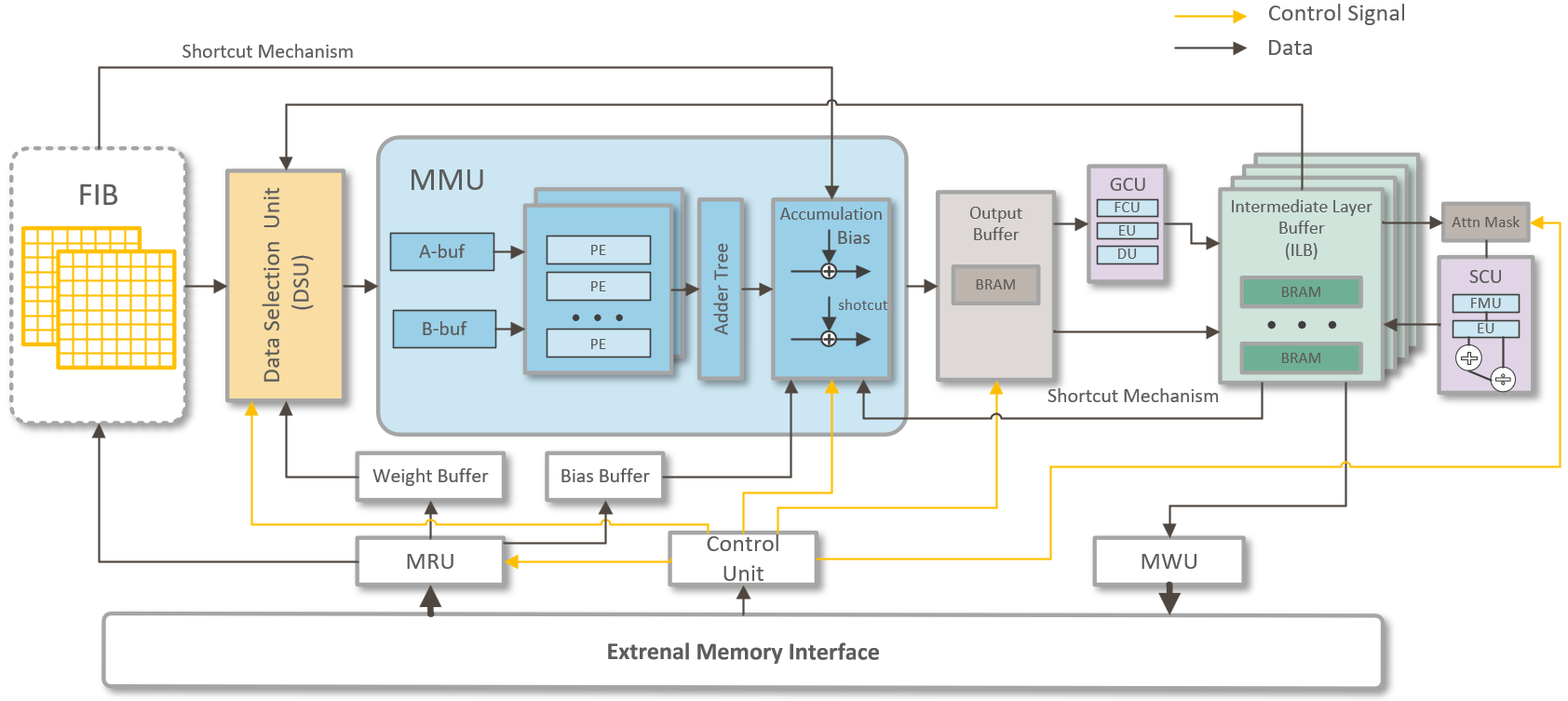}
\caption{The overall architecture of the proposed accelerator.}
\label{swin_acc}
\end{figure*}

\subsection{Accelerator Overall Architecture}
 
Fig. \ref{swin_acc}. illustrates the overall architecture of the proposed accelerator. The accelerator consists of a memory read unit (MRU), a memory write unit (MWU), a feature map input buffer(FIB), a data selection unit (DSU), a matrix multiplication unit (MMU), intermediate layer buffer (ILB), a GELU computation unit (GCU), a Softmax computation unit (SCU), and a control unit. It also includes additional buffers such as weight buffer, bias buffer, and output buffer. (Please note that ILB is not a single buffer, but rather a collection of intermediate layer buffers.)

The accelerator has three operational modes, corresponding to the computations of Patch Embedding, Patch Merging, and the Swin Transformer Block, respectively. In the mode of computations of Patch Embedding, MRU is responsible for reading data from external memory through the External Memory Interface and writing it into the FIB, DSU sending input image data and weight parameters to the MMU for convolutional computations. The output buffer writes the convolutional calculation results into the ILB, and then the MWU transfers feature image data to external memory storage. The dataflow of the Patch Merging computations mode is similar to the mode Patch Embedding. We will focus on explaining the dataflow of the Swin Transformer Block.

The FPGA accelerator we have designed is capable of executing the entire computation of the Swin Transformer Block in a single round, which includes operations like Shifted Window Self-Attention (SWMSA) or Window-based Multi-head Self-Attention (WMSA), Shortcut Mechanism, and computations of the feedforward neural network (FFN). SWMSA or WMSA is the central component of the Swin Transformer Block. First, the MRU reads the feature map data and weight parameters of the partitioned windows from the external memory through the External Memory Interface, guided by the control signals from the control unit. The DSU takes the feature maps from the partitioned windows in the FIB and the corresponding weight parameters from the weight buffer, sending them to the MMU for linear computations. This process generates the three essential feature vectors (QKV) needed for multi-head self-attention (Multiply the weight parameters corresponding to Q by a scaling factor to avoid performing a scaling operation in attention computation). Subsequently, the output buffer writes the computed QKV feature vectors from the MMU into the corresponding buffers within the ILB. The DSU reads the Head-splited Q and K matrices from the ILB and computes the matrix multiplication of Q and the transposed K matrices using the MMU, resulting in attention scores. These scores are then subjected to Softmax computation within the SCU to obtain attention weights (In SWMSA, we need to apply to mask of window attention). Next, the DSU reads the attention weights and the V feature vectors and performs matrix multiplication using the MMU to obtain the final self-attention representation. The resulting self-attention representation is further mapped to a new feature space through linear transformations. Add a Shortcut Mechanism by directly establishing an addition operation between the Accumulation Module within the MMU and FIB, enabling input and output summation. Finally, write the computed result into the ILB.

The dataflow of the FFN is relatively straightforward. The MMU performs a linear transformation on the input, expanding or scaling its feature dimensions by a factor of $M_r$. The output from the feature dimension expansion is then processed through the GCU to compute the GELU activation function. The output from the feature dimension scaling is combined with the input to establish a Shortcut Mechanism, achieving an addition between the input and output. The computed result of the FFN is written into external memory through the External Memory Interface using the MWU.

\subsection{Matrix Multiplication Unit (MMU)}

The proposed accelerator adopts a single matrix multiply engine architecture. As previously mentioned, in our accelerator, all the linear computations, including convolutions and matrix multiplications, in Patch Embedding, Patch Merging, WMSA, SWMSA, and FFN utilize the same MMU. Meanwhile, In the process of using the Swin Transformer model for image processing, over 95$\%$ of the operations are linear computations. These linear computations include convolution operations for Patch Embedding, linear transformations for Patch Merging, linear transformations for generating the query (Q), key (K), and value (V) feature vectors, dot product between query vector (Q) and key vector (K) to obtain similarity matrices, weighted summation of attention weight matrix and feature vectors (V), linear computations in the projection layer, and linear computations in the two fully connected layers of the FFN. The efficiency of handling these linear computations will directly determine the efficiency of the accelerator. 

Inspired by \cite{khan2021npe} and \cite{lit2022auto}, we propose a novel, versatile, and efficient FPGA MMU. This MMU is designed for computing convolutions in Patch Embedding, matrix multiplications in Swin Transformer Block, and Patch Merging. We also introduce data selection methods and output accumulation methods for each linear computation, enabling the utilization of the MMU for computing all the aforementioned linear functions. 

The core idea behind our MMU design is to employ blocked matrix multiplication for computing matrix products. The MMU simultaneously takes an input A of shape $M^2\times c_i$ and an input B of shape $c_i\times c_o$. The MMU consists of 32 PEs, with each PE comprising 49 parallel multiplication units. Each PE is responsible for the multiplication computation between an $M^2\times c_i$ matrix and a $c_i\times1$ vector, followed by parallel addition in the adder tree. In FPGA, each DSP48E1 is responsible for the multiplication of a pair of 16-bit fixed-point numbers, and the entire MMU requires 1568 DSP compute units. The MMU writes the computed result, a matrix with a shape of $M^2\times c_o$, into an accumulation buffer. Accumulation is performed based on different linear functions.

In the Swin Transformer block, the linear computations can all be regarded as matrix multiplications. In the Swin Transformer Block, the query vector (Q) key vector (K), and value vector (V) are divided into multiple heads, each with a dimension of 32. Meanwhile, in other matrix multiplication calculations, the dimensions of the weight matrices and value vectors usually remain multiples of 32. This is also the reason why we set the dimension $c_o$ as 32. However, there is an exception during the computation of attention scores, where the dimension of the transposed key vector (often referring to the transposed K or a sub-matrix of K) might not be a multiple of 32. 

\begin{figure}[!t]
\centering
\includegraphics[width=3.4in]{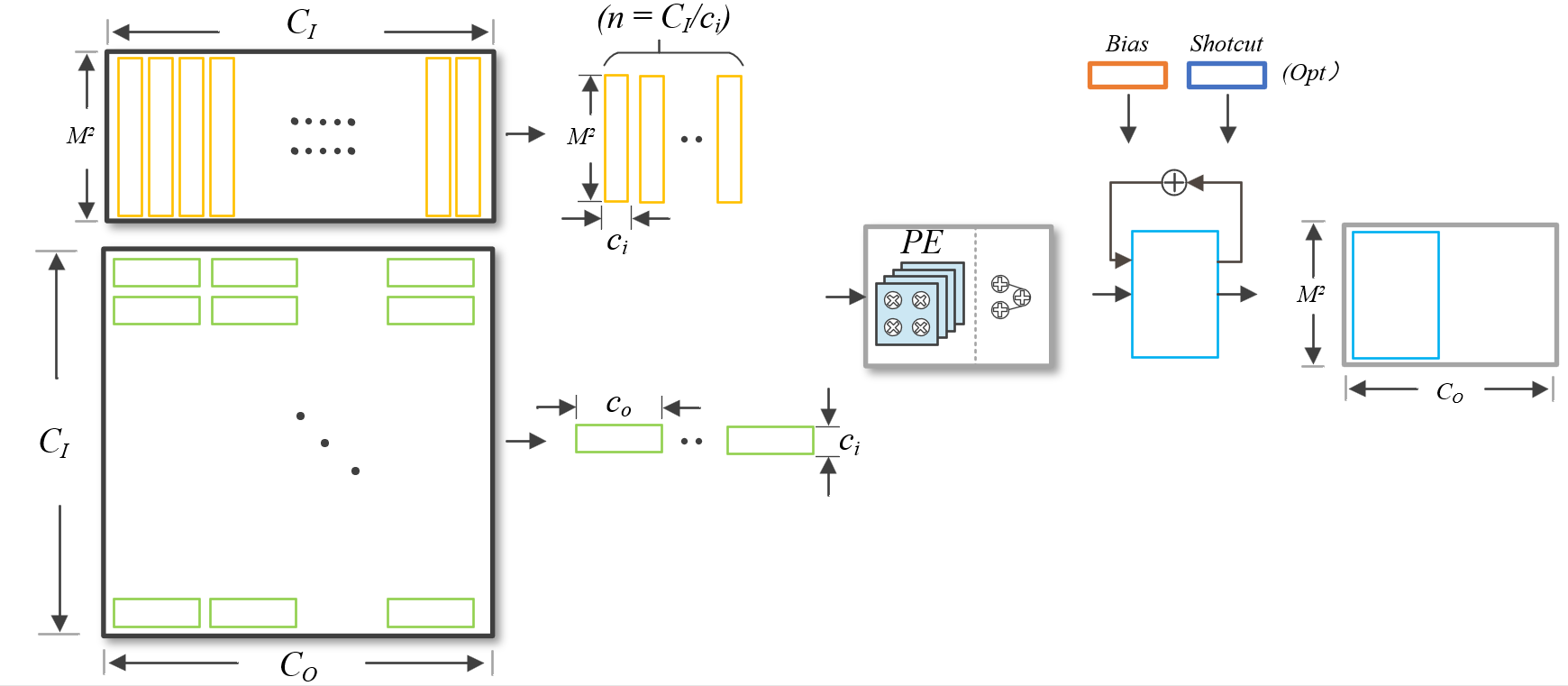}
\caption{Tiling method for Matrix Multiplication. In the swin transformer architecture, the number of rows in the input matrix A for all matrix multiplications is equal to the square of the window size $M^2$. Matrix computations are performed using blocked matrix multiplication, with each block requiring $C_I/c_i$ computation cycles to complete.}
\label{Linear_cpu}
\end{figure}

\begin{figure}[!t]
\centering
\includegraphics[width=3.4in]{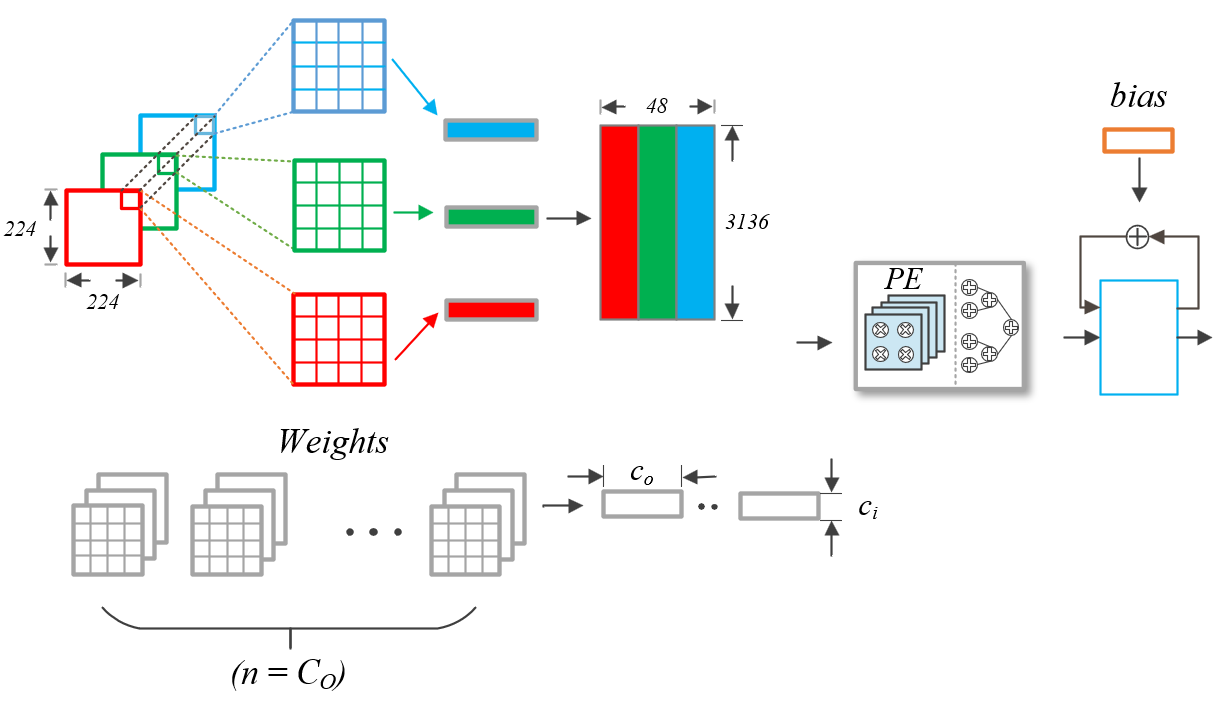}
\caption{Tiling and flattening method for Patch Embedding. The weight parameters are flattened in the same way as the input image}
\label{patch_embed}
\end{figure}

\begin{table}[htb]
\caption{Notations for matrix multiplication processing.\label{tab:table0}}
\centering
\begin{tabular}{c|c}
\toprule
Notation & Description \\
\midrule
$X$ & Input matrix  \\
$M$ & Each window size for window partition \\
$c_i$ & Number of dimensions for data in input channel in MMU \\
$c_o$ & Number of dimensions for data in output channel in MMU \\
$C_I$ & Number of input channels \\
$C_O$ & Number of output channels \\
\bottomrule
\end{tabular}
\end{table}

Taking the linear transformation for generating QKV feature vectors as an example, the input feature matrix X (with a shape of $M^2\times C_I$) is subjected to matrix multiplications with the weight parameter matrices $WQ$, $WK$, and $WV$ (all having shapes of $C_I\times C_O$) respectively. This process yields corresponding Q, K, and V vectors (all with shapes of $M^2\times C_O$). Matrix multiplication is performed using the loop tiling shown in Fig. \ref{Linear_cpu}, where each input feature matrix X, weight parameter matrices, and output data are segmented into tiles. Through pipelining and loop unrolling, processing elements (PEs) can concurrently handle $M^2\times c_i\times c_o$ multiplicative operations. The matrix ($M^2\times c_o$) obtained from PEs' calculations is subjected to accumulation in the Accumulation module over ($C_I/c_o$) cycles, ultimately writing the output vector into the output buffer. The aforementioned approach is applicable to all linear function computations within the Swin Transformer block.

However, there is an exception. In the Swin Transformer, when performing matrix multiplication between the transposed query vector (Q) and key vector (K) after the head-splitting, using the aforementioned approach is not suitable due to the fact that the dimension of the key vector (K) is not divisible by $c_o$(32). To address this issue, we propose a method where, when the input cannot completely fill the buffer B. We expand the $K^T$ matrix and pad the remaining elements with zeros to meet the input requirements of the MMU. This approach undoubtedly leads to invalid computation and wastage of FPGA resources such as DSP blocks and lookup tables (LUTs). However, invalid computations account for only 1.2$\%$ of all linear computations in Swin Transformer Block (specific calculations will be elaborated in section \ref{ANALYSIS AND EVALUATION}). Given the universality and efficiency of the MMU, this resource wastage is negligible.

In the Patch Embedding module of the Swin Transformer, a convolutional operation with a window size of 4x4 and a stride of 4 is applied to segment the input image into image patches. Specifically, considering an input image of size 224x224, the image can be divided into patches of size 4x4. Each patch is then flattened into a vector of length 16. These vectors are arranged in sequence to form a matrix X, with $\left(\frac{224-4}{4}+1\right)\times\ \left(\frac{224-4}{4}+1\right)=3136$ rows and 16 columns. Next, this matrix X is multiplied with a suitable weight matrix W to obtain a feature matrix.

\subsection{The Architecture of SCU (Softmax Compute Unit)}

\subsubsection{The Overall Architecture for SCU (Softmax Compute Unit)}

\begin{figure}[!t]
\centering
\includegraphics[width=3.0in]{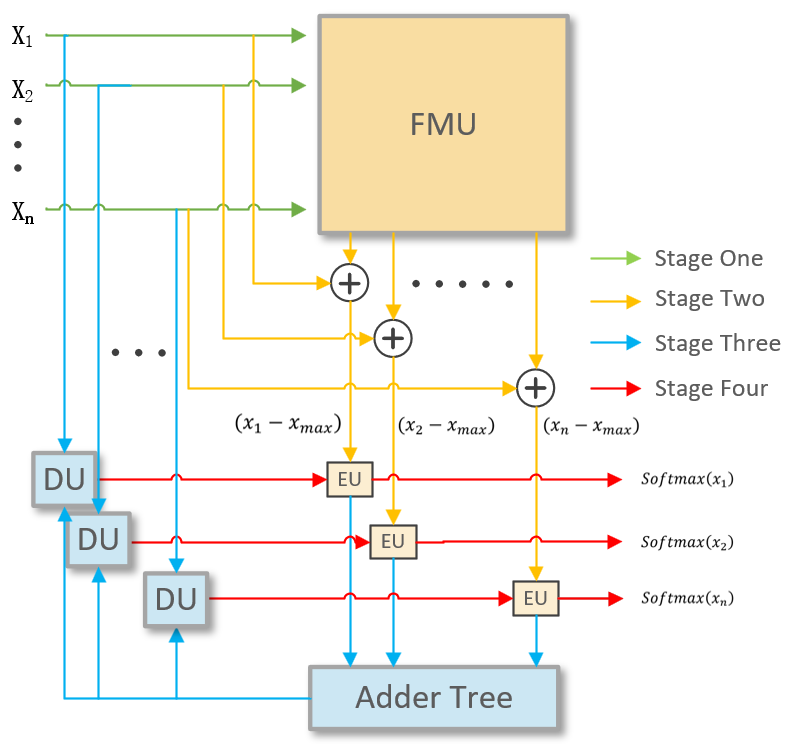}
\caption{The Structure of SCU. To ensure computational efficiency, the parallelism of both the EU and DU is set equal to the number of input data.}
\label{SCU}
\end{figure}

As shown in Fig. \ref{SCU}, the SCU consists of several submodules, including a Find Max Unit (FMU), Exponential Unit (EU), Adder Tree, and a Division Exponent Calculation Unit (DU). The computation process of Softmax can be divided into four stages in its dataflow. Stage One: The FMU receives input data and identifies the maximum value $x_{max}$ within a set of data. Stage Two: The input data is subtracted from the output of FMU to obtain $x_i-\ x_{max}$, which is then fed into the EU for exponential computation. Stage Three: The Adder Tree calculates the cumulative sum of the EU outputs, and the DU receives the cumulative sum from the Adder Tree to compute the exponent of Formula (11) along with the input data. Stage Four: The EU processes the output from DU and computes the final result of the Softmax. 

\subsubsection{The Structure of FMU (Find Max Unit)}

\begin{figure}[!t]
\centering
\includegraphics[width=3.0in]{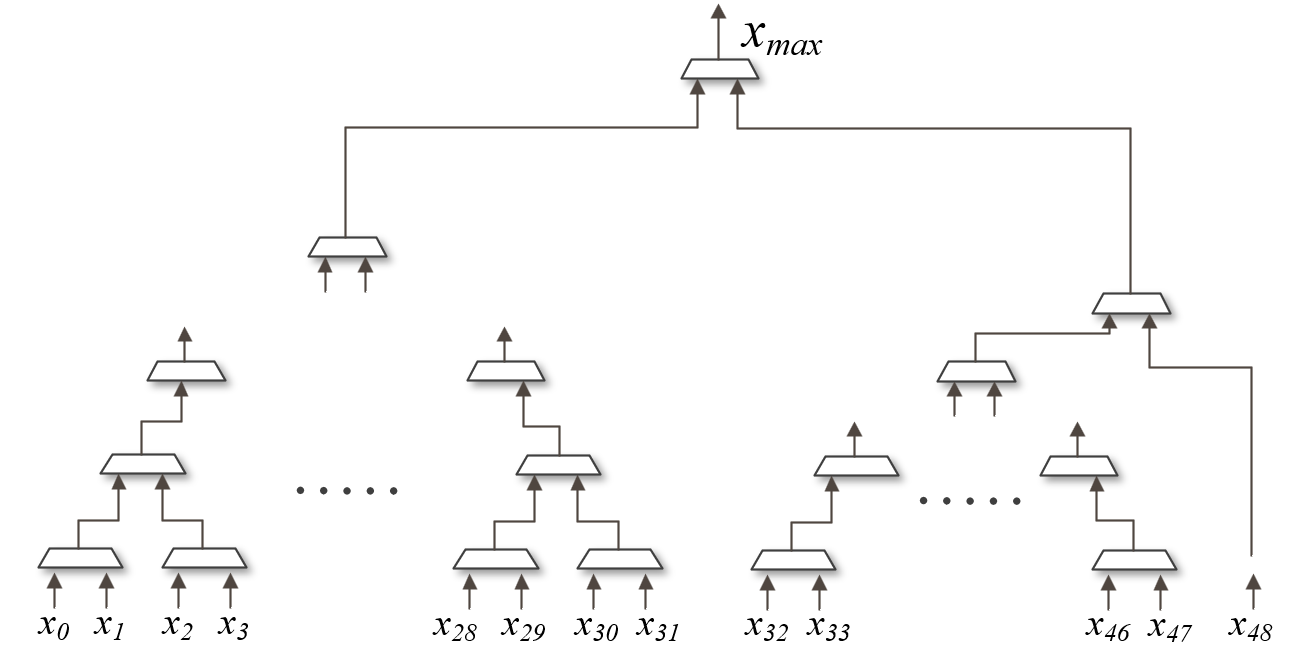}
\caption{The Structure of FMU. Due to the difference in the number of elements in the groups, Group 2 completes its operation faster than Group 1. Once Group 2 finishes finding the maximum value, the result is compared to $x_{48}$ in order to further reduce an additional cycle in the execution.}
\label{FMU}
\end{figure}

In general, identifying the maximum value within a set of input data can be achieved using a traversal method, which is the simplest and most straightforward approach. However, this method results in a time complexity of $O(n)$ for the operation. After obtaining the similarity matrix from the QK dot product, the shape of the similarity matrix is 49x49. The operation of pairwise comparison to traverse all elements requires 48 clock cycles.  For an efficient accelerator, such processing speed is unacceptable. We propose an FMU that employs a parallelization approach for efficiently finding the maximum element within a set of vectors. The structure of the FMU is illustrated in Fig. \ref{FMU}. The elements in our vector are divided into multiple groups, where the number of elements in each group satisfies three conditions: 1. The number of elements in each group is a power of two. 2. The number of elements in each group is maximized as much as possible. 3. The sum of the element counts in all groups is equal to the total number of elements in the vector. As shown in Fig. \ref{FMU}, we categorize $(x_0,\ x_1,\ \ldots,x_{31})$ as Group One, $(x_{32},\ x_{33},\ \ldots,x_{47})$ as Group Two, and $(x_{48})$ as Group Three. Taking Group 1 as an example, the 32 elements in Group 1 are divided into 16 pairs, with each pair containing 2 elements ($x_0$ and $x_1$), making a total of 16 pairs. Within a single clock cycle, comparisons are performed for each pair, resulting in 16 larger elements. In the subsequent clock cycle, these 16 larger elements are compared in pairs again, yielding 8 even larger elements. This process continues iteratively, with the number of elements halving in each clock cycle, until a single largest element is eventually obtained. The same approach is employed for finding the maximum value in other groups as well. The time complexity of the proposed method for finding the maximum value is $O\left(\left\lceil\log _2 n\right\rceil\right)$. In theory, finding the maximum value of elements in a vector of length 49 would require 6 cycles. This represents a significant improvement in speed compared to the previous 48 clock cycles.

\subsubsection{The Structure of EU (Execution Unit)}

\begin{figure}[!t]
\centering
\includegraphics[width=3.0in]{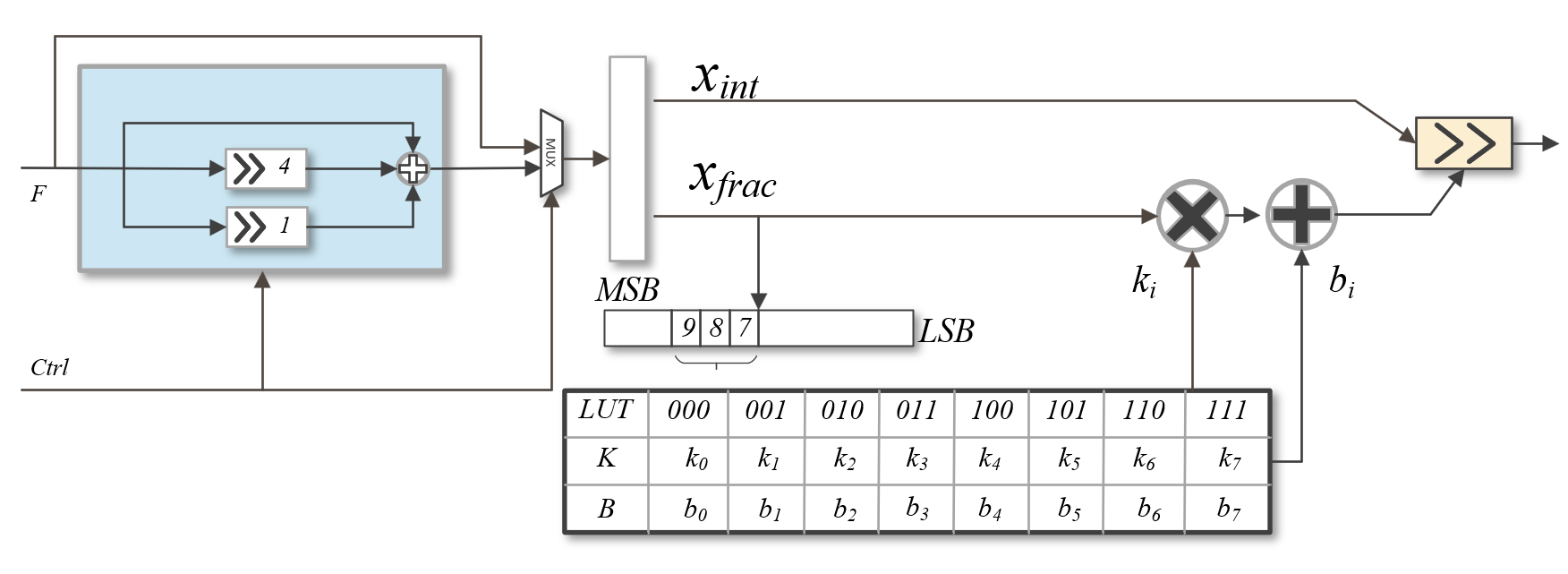}
\caption{The Structure of EU. In stage 2, input $F$ is multiplied by $log _2 e $ through addition and shift operations. In stage 4, the input does not do this.}
\label{EU}
\end{figure}

The structure of the EU (Execution Unit) is illustrated in Fig. \ref{EU}. The input control signal determines whether the input data undergoes exponential calculations with a base of 2 or a base of e. According to the formula (10), we use the 9th, 8th, and 7th bits of $frac\left(x_i\right)$ as flags, and based on the values of $Ki$ and $Bi$ stored in the LUT, perform piecewise linear approximation to calculate $2^{frac{\left(x_i\right)}}$.

\subsubsection{The structure of the DU (Division Exponent Calculation Unit)}

\begin{figure}[!t]
\centering
\includegraphics[width=3.0in]{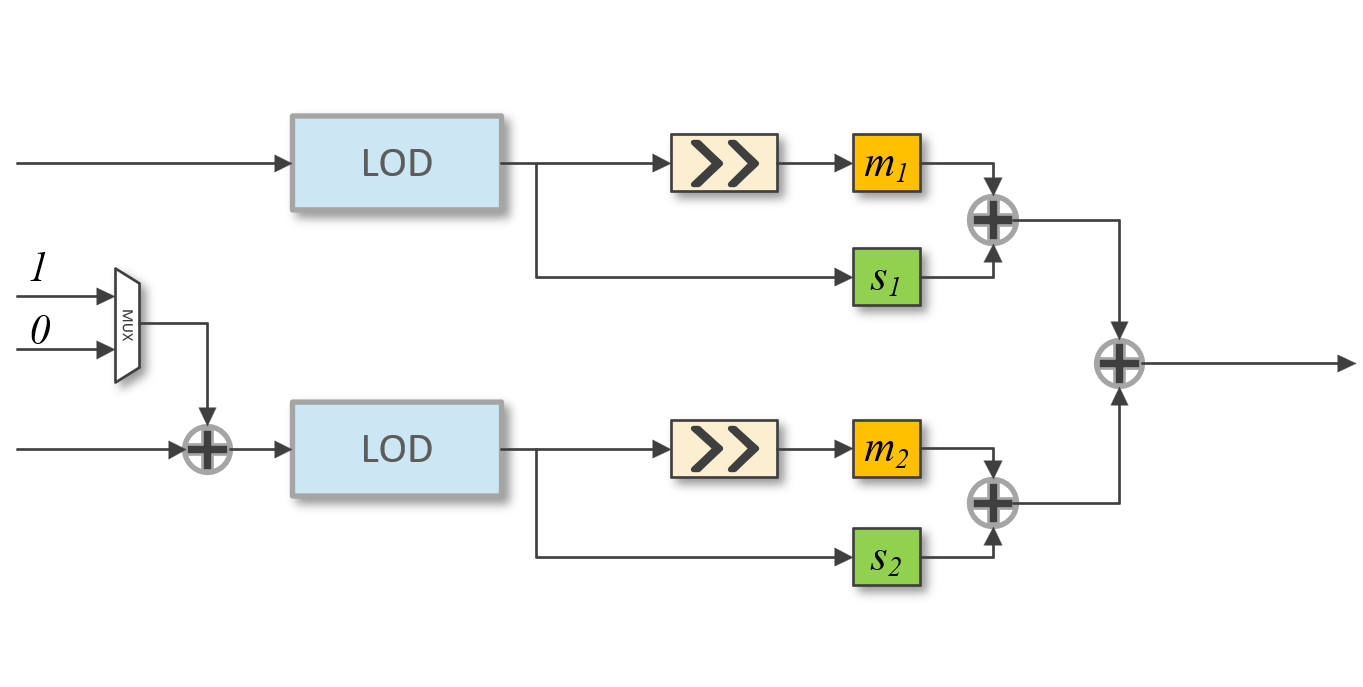}
\caption{The Structure of DU.  The structure of DU in GCU is identical to that in SCU. The only difference is whether DUs should add 1 to the denominator inputs based on the denominators of formulas (6) and (8).}
\label{DU}
\end{figure}

The DU receives two inputs. The highest position of bit 1 in F is determined using an A Leading One Detector (LOD). The values of m and w can be easily calculated. Finally, the exponent in the formula (12) is computed based on the values of $m1$, $m2$, $w1$, and $w2$.

\subsection{The Overall Architecture of GCU(GELU Compute Unit)}

\begin{figure}[htb]
\centering
\includegraphics[width=3.2in]{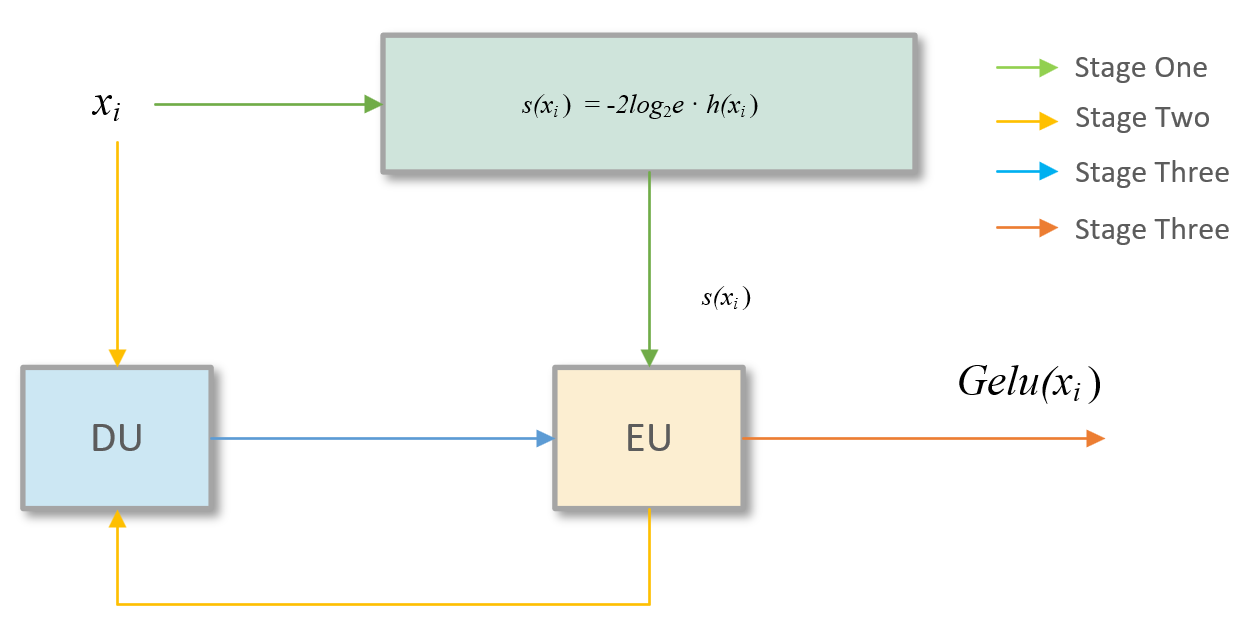}
\caption{The Structure of GCU. According to formula (9) and the information provided in section B, in polynomial calculations, except for the computation of $x^3_i$, all other calculations are accomplished through shift and addition operations.}
\label{GCU}
\end{figure}

The structure of GCU is relatively simple, consisting of a polynomial computation unit, an EU, and a GU. The computation process of GCU can be divided into four stages in its dataflow. Stage one: GCU receives input $x_i$ and completes polynomial computation $s\left(x_i\right)=-2 \log _2 e \cdot \sqrt{\frac{2}{\pi}}\left(x_i+0.044715 x_i^3\right)$. Stage two: The EU receives the output from the polynomial computation unit and calculates $2^{-2 \log _2 e \cdot h\left(x_i\right)}$. Stage three: The DU computes the exponent of Formula (12). Stage four: The EU processes the output from the DU and calculates the final result of the Gelu.

\section{ANALYSIS AND EVALUATION}\label{ANALYSIS AND EVALUATION}

\subsection{Analysis of Invalid Computations}

In the preceding section, the process of computing attention scores within the scaled dot-product attention was explained. To meet the input requirements of the MMU, we expand the $K^T$ matrix and pad the remaining elements with zeros. This approach undoubtedly leads to invalid calculations and wastes resources such as DSPs and lookup tables (LUTs). We will analyze the proportion of invalid calculations caused by this method to the total linear calculation.

In Swin Transformer block, the computational complexity of \textbf{W-MSA} or \textbf{SW-MSA} can be represented as follows:

\begin{equation}
\label{deqn_ex13}
\Omega(W-M S A)=4 h w C^2+2 M^2 h w C
\end{equation}

With the FFN expansion ratio factor $M_r$ of 4, the computational complexity of the FFN can be expressed as follows:

\begin{equation}
\label{deqn_ex14}
\Omega(F F N)=8 h w C^2
\end{equation}

The computational complexity of the dot product between Q and K can be expressed as:

\begin{equation}
\label{deqn_ex15}
\Omega\left(q \cdot k^T\right)=M^2 h w C
\end{equation}

After expanding the $K^T$ matrix, the computational complexity of matrix multiplication can be expressed as:

\begin{equation}
\label{deqn_ex16}
\Omega\left(q \cdot k^T_e\right)=2 c_o h w C
\end{equation}

Therefore, in the linear computation of the proposed accelerator's Swin Transformer block, the proportion of invalid computations to the total linear computation, represented by U, can be indicated by the formula (17):

\begin{equation}
\label{deqn_ex17}
U=\frac{2 c_o h w C-M^2 h w C}{12 h w C^2+2 M^2 h w C}
\end{equation}

In Swin-T, Swin-S, and Swin-B models, the Depth of layer is $<2, 2, 6, 2>$, $<2, 2, 18, 2>$ and $<2, 2, 18, 2>$ respectively. The input image size is 224x224, and the channel numbers \textbf{C} are 96, 96, and 128 respectively. The window size is $M\times M=7\times7$. According to the formula (17), we can calculate that U is 1.2$\%$. Therefore, we can conclude that the proportion of invalid calculations and wasted DSP resources is negligible.

\subsection{Verification of The Feasibility of Replacing LN by BN}

First, we replicated the validation of \cite{yao2021leveraging}'s work and assessed the feasibility of replacing Batch Normalization (BN) with Layer Normalization (LN) across three scales of Swin Transformers, namely Swin-T, Swin-S, and Swin-B. We utilized the ImageNet-1K dataset as the benchmark training set.

During our training process, we employed the AdamW optimizer for training over 300 epochs. Within this period, we conducted a linear warm-up for 20 epochs and adopted a cosine decay learning rate scheduler. Our chosen batch size was 1024, with an initial learning rate of 0.001, and a weight decay of 0.05 was applied. Across the three model variants, Swin-T, Swin-S, and Swin-B, we consistently used the default input image resolution of 224. To ensure efficient training, these models were deployed on 8 Nvidia GeForce RTX 4090 GPUs. The training results are shown in Table \ref{tab:table1}. The Baseline accuracy decreased by 0.6$\%$, 0.3$\%$, and 0.7$\%$, respectively, compared with the LN normalization scheme. Which shows an acceptable result.

 \begin{table}[htb]
\caption{Verification of the feasibility of replacing LN by BN on ImageNet-1K classification.\label{tab:table1}}
\centering
\begin{tabular}{cccc}
\toprule
Model & Baseline(LN) & \cite{yao2021leveraging}(BN) & Ours\\
\midrule
Swin-T(Top-1 acc) & 81.3$\%$ & 80.9$\%$ & 80.7$\%$ (0.6$\%$ $\downarrow$) \\
Swin-S(Top-1 acc) & 83.0$\%$ & 82.8$\%$ & 82.7$\%$ (0.3$\%$ $\downarrow$)\\
Swin-B(Top-1 acc) & 85.5$\%$ & 83.1$\%$ & 82.8$\%$ (0.7$\%$ $\downarrow$)\\
\bottomrule
\end{tabular}
\end{table}

\subsection{Quantification Methods}

In \cite{lit2022auto}, the authors employed techniques named POT quantization for multi-scale ViT quantization. They demonstrated the feasibility of using 8-bit and even 4-bit quantization in ViT models. In \cite{li2023high} the authors proposed a high-speed reconfigurable architecture that utilizes 16-bit fixed-point representation for running all Softmax and GELU operations. In the work \cite{zhu2020efficient}, a precision-adjustable architecture for the Softmax function was developed, with all inputs and outputs represented in 16-bit format, achieving both efficiency and adjustability. Furthermore, \cite{vasyltsov2021efficient} even explored the use of 8-bit quantization for Softmax function computations, achieving minimal precision loss while working with attention mechanisms in deep neural networks.

The proposed accelerator in this study adopts a full-quantized method, eliminating a significant number of intermediate quantized and de-quantized operations. This approach also extends to quantizing biases, leading to an inevitable loss of a certain degree of flexibility. However, throughout this process, there is no involvement of float32 computations. This design choice enhances hardware efficiency, particularly for hardware like FPGA that may not be as well-suited to handling float operations. In order to maintain accuracy, the matrix multiplication can be quantized to 16-bit fixed point without any noticeable loss in precision.

Hence, to ensure both accuracy and efficiency, the proposed accelerator in this paper utilizes 16-bit fixed-point computation for linear functions and 16-bit fixed-point calculations for Softmax and Gelu nonlinear functions.

\subsection{Implementation of Accelerator}
In order to evaluate the proposed design, the accelerator was implemented on a Xilinx XCZU19EG FPGA, which includes 522.7K LUTs, 1968 DSPs, and 984 BRAMs. The high-level (C/C++) based designs of the accelerator are synthesized and generated using Xilinx Vitis 2022.2, and then we use Xilinx Vivado 2022.2 to synthesize and implement the complete project on the selected FPGA.

\subsubsection{Resource Utilization of Submodules}
In Table \ref{tab:table2}, we present the consumption of DSPs, LUTs, FFs, and BRAMs by submodules such as MMU, SCU, and GCU. MMU accounts for the majority of DSP utilization.

\begin{table}[!t]
\caption{Resource Utilization of Submodule\label{tab:table2}}
\centering
\begin{tabular}{c|cccc}
\hline
Submodule & DSP & LUT & FF & BRAM\\
\hline
MMU & 1568(79.7$\%$) & 198960(38.0$\%$) & 14115 & 14\\
SCU & 49(2.5$\%$) & 41184(7.9$\%$) & 18708 & 4\\
GCU & 98(5.0$\%$) & 53482(10.2$\%$) & 5745 & 4\\
\hline
\end{tabular}
\end{table}

\subsubsection{Resource Utilization of the Accelerators}
In Table \ref{tab:table3}, we present the resource utilization of accelerators for three scales of Swin Transformers, namely Swin-T, Swin-S, and Swin-B.

\begin{table}[!t]
\caption{Resource Utilization of the Accelerators\label{tab:table3}}
\centering
\begin{tabular}{c|cccc}
\hline
Model & DSP & LUT & FF & BRAM\\
\hline
Swin-T & 1727(87.8$\%$) & 434(83.1$\%$) & 271(25.9$\%$) & 244(25.2$\%$)\\
Swin-S & 1727(87.8$\%$) & 434(83.1$\%$) & 271(25.9$\%$) & 244(25.2$\%$)\\
Swin-B & 1733(88.0$\%$) & 451(86.4$\%$) & 378(36.2$\%$) & 338(34.9$\%$)\\
\hline
\end{tabular}
\end{table}

\subsection{Comparison with CPU, GPU, and FPGA}
In this section, we will compare the performance of the proposed FPGA-based accelerator in terms of Relative Speedup and power consumption with that of CPU and GPU. We use Pytorch (v2.0.0) frameworks on the WSL2 Ubuntu 20.04 operating system with CUDA 11.8 to run model inference on both CPU and GPU. CPU power consumption is monitored by HWiNFO64 CPU Package Power, and GPU power consumption is monitored by MSI Afterburner. The FPGA Power consumption is estimated by Vivado Report Power. The detailed information of CPU and CPU is listed as follows.

\begin{enumerate}{}{}
\item{CPU Baseline: AMD Ryzen 5700X with 8 physical cores, 12 threads. Operating at 4.0HHz.}
\item{GPU Baseline: Nvidia GeForce RTX 2080 Ti with 11-GB GDDR6 operating at 1499 MHz.}
\end{enumerate}

\begin{figure}[!t]
\centering
\includegraphics[width=3.0in]{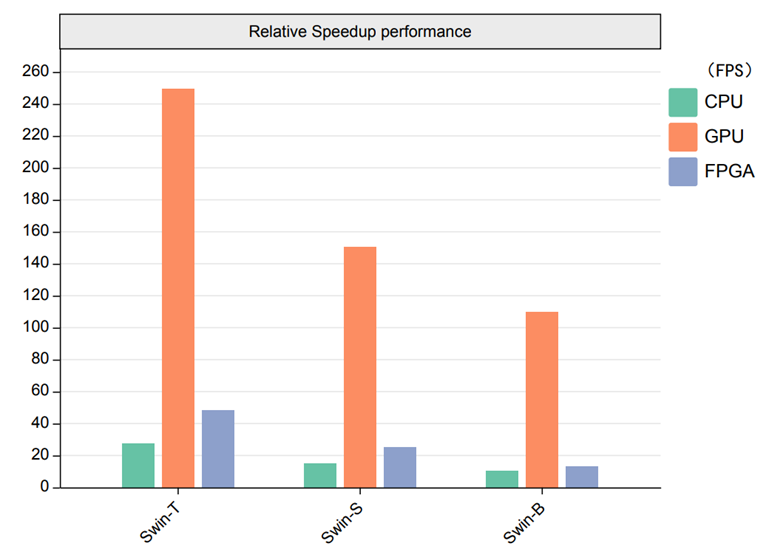}
\caption{Relative speedup for CPU, GPU, and Accelerator. The accelerator's speedup surpasses the performance of the CPU. Although the accelerator's speedup may not match that of a GPU, our design's power efficiency is superior, making it well-suited for computationally constrained situations like edge computing applications.}
\label{speedup}
\end{figure}

\begin{figure}[!t]
\centering
\includegraphics[width=3.0in]{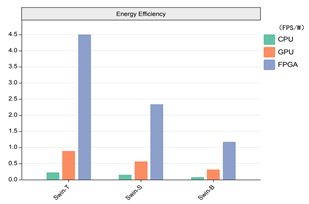}
\caption{Energy Efficiency for CPU, GPU, and Accelerator. Energy Efficiency represents the ratio of FPS (Frames Per Second) to power consumption. During inference, the approximate power consumption of the CPU is 120W, while the GPU consumes approximately 240W. Meanwhile, the accelerator's power consumption is only around 10W.}
\label{Energy eff}
\end{figure}

\begin{table*}[htb]
\caption{Comparison with Related Swin Transformer Accelerators\label{tab:table4}}
\centering
\begin{tabular}{c|cccccccc}
\toprule
Design & Model & Platform & Frequency (MHz) & Precision & Power  & Frame Rate  & Throughput  & DSPs $\downarrow$\\
& & & & & (W) $\downarrow$ & (FPS) $\uparrow$ & (GOPS) $\uparrow$ &
\\
\midrule
\cite{wang2022via} & Swin-T & Alveo U50 & 300 & Float16 & 39 & * & 309.6 & 2420\\
\cite{nag2023vita} & Swin-T & XC7Z020 & 150 & Fix8 & 0.88 & 8.71 & * & *\\
\cite{hu2022hardware} & Window Attention & ZCU102 & 100& Fix8 & * & * & 75.17 & 70\\
\textbf{Ours} & Swin-T & XCZU19EG & 200 & Fix16 & 10.69 & 48.1 & 431.2 & 1727\\
\textbf{Ours} & Swin-S & XCZU19EG & 200 & Fix16 & 10.69 & 25.0 & 436.4 & 1727\\
\textbf{Ours} & Swin-B & XCZU19EG & 200 & Fix16 & 11.11 & 13.1 & 403.5 & 1733\\
\bottomrule
\end{tabular}
\end{table*}

Each device’s relative speedup performance showed in Fig. \ref{speedup}, and Energy Efficiency showed in Fig. \ref{Energy eff}.

From Fig. \ref{speedup}, compared to the CPU, our accelerator achieves approximately 1.76x, 1.66x, and 1.25x speedup on Swin-T, Swin-S, and Swin-B models respectively. Meanwhile, compared with CPU, we achieved approximately 20.45x, 18.60x, and 14.63x energy efficiency (FPS/power consumption) improvement on Swin-T, Swin-S, and Swin-B models, respectively. Compared to the GPU, our accelerator achieved approximately 0.20x, 0.17x, and 0.12x speedup on Swin-T, Swin-S, and Swin-B models, respectively. Meanwhile, compared with GPU, we achieved approximately 5.05x, 4.42x, and 3.00x energy efficiency improvement on Swin-T, Swin-S, and Swin-B models, respectively.

\subsection{Comparison with Related Works}
To further evaluate the performance of our Swin Transformer accelerator, we compare our accelerator with related existing accelerators. The several accelerator performances of details shown in Table \ref{tab:table4}(\cite{hu2022hardware} did not implement a complete Swin Transformer, but rather implemented a sub-module called Window Attention in Swin Transformer).

Compared to \cite{wang2022via}, we not only used fewer DSP resources, but also achieved approximately 0.25x power consumption, and achieved approximately 1.40x throughput. Achieved lower power consumption as well as higher throughput. Despite \cite{nag2023vita} has proposed a more energy-efficient accelerator, compared to \cite{nag2023vita}, Our accelerator achieved a higher frame rate, which is 5.52x that of \cite{nag2023vita}. \cite{hu2022hardware} implement  Windows Attention rather than a complete Swin Transformer model on FPGA, we still achieved 5.80x throughput Compared with \cite{hu2022hardware}. As far as our current knowledge goes, the accelerator we proposed is the fastest FPGA-based accelerator for Swin Transformer.

\section{CONCLUSION}\label{section}

In this paper, we proposed an efficient FPGA-based accelerator designed for the Swin Transformer. We modified the model architecture of the Swin Transformer by substituting Layer Normalization (LN) with Batch Normalization (BN), allowing the fusion of BN with linear layers for efficient inference. The modified Swin Transformer achieves a minor accuracy reduction on ImageNet-1k. We designed an efficient MMU with related accelerator optimizations and proposed hardware-friendly architecture for Softmax Function and GELU Function. Our experiments proved that our proposed accelerator respectively achieves 1.76x, 1.66x, and 1.25x speedup and 20.45x, 18.60x, and 14.63x energy efficiency compared to AMD Ryzen 5700X CPU. Compared with a far more computationally intensive middle-end GPU Nvidia GeForce RTX 2080 Ti, our design has achieved 5.05x, 4.42x, and 3.00x energy efficiency, which makes it a suitable choice for edge computation scenarios.
In the meanwhile, compared with current state-of-the-art accelerators, our accelerator has a better speedup performance and a better energy efficiency performance.

\bibliographystyle{IEEEtran}
\bibliography{sample}

\end{document}